\documentstyle[aps,prl,epsfig,twocolumn,floats,amssymb,amsfonts]{revtex}
\begin{document}
\bibliographystyle{prsty}
\global\firstfigfalse
\global\firsttabfalse
\title{Spin configurations of carbon nanotube in a nonuniform  external
        potential}
\author{Yuval Oreg, Krzysztof Byczuk \cite{NT:KB} and Bertrand I. Halperin}
\address{ Lyman Laboratory of Physics, Harvard University, Cambridge,
  MA 02138 \\ {\rm \today} \\ \bigskip \parbox{14cm} 
  {\rm We study, theoretically, the ground state spin of a carbon
    nanotube in the presence of an external potential. We find that
    when the external potential is applied to a part of the nanotube,
    its variation changes the single electron spectrum significantly.
    This, in combination with Coulomb repulsion and the symmetry
    properties of a finite length armchair nanotube induces spin flips
    in the ground state when the external potential is varied. We
    discuss the possible application of our theory to recent measurements of
    Coulomb blocked peaks and their dependence on a weak magnetic field
    in armchair carbon nanotubes.  \smallskip \\ PACS numbers
    61.48.+c, 73.23.-b, 71.10.-w, 71.24.+q, 75.10.Lp\vspace{-0.5cm}
} } \maketitle

Carbon nanotubes have attracted continuous attention since their
discovery \cite{NT:Iijima91}.  Their unique mechanical and electrical
properties enable to design and build several systems, some may be
used for practical applications.  When they are rolled up in a certain
way the electron states near the Fermi level have a one-dimensional
(1D) character\cite{NT:Few}. The Coulomb interaction between the
electrons in low-dimensional systems, especially in 1D, affects
significantly the physical properties of the electrons.  Indeed, power
law behavior of the tunneling conductance was found
experimentally\cite{NT:Bockrath99}.  Recently, Coulomb blockade peaks
were investigated in a single-electron transport experiments through
short ($\sim 0.2- 3 \;\mu{\rm m}$) isolated metallic
nanotubes\cite{NT:Bockroth97,NT:Tans97,NT:Bezryadin98,NT:Tans98,NT:Cobden98,NT:Postma99}.

In the experiments of Tans et al. \cite{NT:Tans98} (I) and Cobden et
al.  \cite{NT:Cobden98} (II) the spin direction of the additional
electrons entering into the tube, as an external gate potential is
varied, was determined various methods.  Within the framework of the
``orthodox theory'', where effects of the Coulomb interaction are
modeled as a constant charging energy, one expects that the electrons
will occupy states in pairs of spin up and spin down.  The last level
is singly or doubly occupied depending on the parity of the total
number of electrons in the system. In that case, successive electrons
entering the nanotube at a low magnetic field should have opposite
spins. This behavior was observed in II, but in I eight successive
electrons entered the tube with parallel spins\cite{NT:Remark99a}.
Shorter parallel spin (PS) sequences were reported in
Ref.~\onlinecite{NT:Postma99}.

The purpose of the present note is to explore one possible mechanism
for PS sequences to occur.  The essential feature is the existence of
two (or more) families of single-electron states in the nanotube,
which respond differently to the applied gate voltage $V_{\rm g}$ and
can cross (or nearly cross) at certain values of $V_{\rm g}$.
Electron-electron interactions can then cause spin flips (i.e.,
internal transitions) which can lead to PS sequences under proper
conditions, as was noted in I.  We point out that this can occur in
ideal defect-free armchair nanotubes, as well as in nanotubes with
sufficient disorder that states at the Fermi energy are localized in
different regions of the tube.

\begin{figure}[h]
  \vglue 0cm \epsfxsize=1\hsize
\epsffile{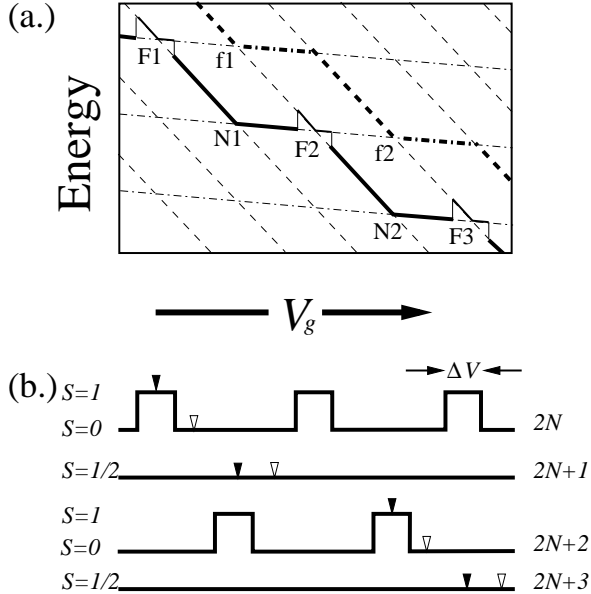}
\caption{\label{fg:levels}
  {\bf (a.)} (Top) Evolution of the discrete single-particle levels of
  an armchair nanotube when a nonuniform electrostatic potential
  $\phi(x)$ on the tube is varied (by changing the voltage, $V_g$, on
  the gate electrode).  The dashed lines represent the (anti-bonding)
  A-band, while the dashed-dot lines represent levels in the (bonding)
  B-band.  The descending solid-line represents the highest occupied
  level for a tube with a fixed even number $2N$ electrons.  Heavy
  portions of the curve show where the highest state is doubly
  occupied ($S=0$), while lighter regions near the crossing points F1,
  F2, F3, show where two states are singly occupied ($S=1$) due to the
  electron-electron interaction.  Similar $S=1$ states will occur for
  $2N+2$ electrons near the crossing points f1 and f2.  For
  odd-electron number, however, such as $2N+3$, no spin-flips occur
  and the highest filled states follows the trajectory indicated by
  the heavy dashed and dot-dashed lines.  Note that $S=1$ states do
  not occur near the crossing points N1 and N2. {\bf(b.)} (Bottom) The
  spin $S$ of the system for fixed numbers of the electrons in the
  system when $V_{\rm g}$ is varied. Electrons added at the black
  triangles would have spins $+1/2$ while those added at the white
  triangles would have alternating $\pm 1/2$ spins.}
\end{figure}

For an infinite armchair nanotube, for a given wavevector parallel to
the tube, there are two energy bands of different symmetry, which we
designate as $B$ and $A$, distinguishing states which are bonding or
anti-bonding for adjacent carbon atoms around the circumference of the
tube. As we show below, levels from the $B$ and $A$ bands respond
differently to a gate potential which is {\em nonuniform} along the
tube axis.  We assume here ideal reflecting boundary conditions at the
ends of the nanotube, which preserve the symmetry, so that the
discrete single particle levels from the two bands can actually cross
each other as the gate potential is varied. (See
Fig.~\ref{fg:levels}.)  If there is {\em weak} mixing between bands on
reflection from a tube end, or due to interactions with the substrate,
near-crossings can still occur, and our conclusions will be unaltered.

Assume now, for example, that there are $2N$ particles in the nanotube
and that the last $B$-level below the Fermi energy is occupied by two
electrons with total spin 0.  As the nonuniform potential (induced by
the gate) is decreased an $A$-level crosses the highest (doubly)
occupied $B$-level at the point $F1$ in Fig.~\ref{fg:levels}a.  In
analogy to the first Hund's rule for partially filled (degenerate)
shell in an atom, the two electrons form a many-body ground state with
a maximum possible spin. They occupy the $A$- and the $B$-levels near
$F1$ point with total spin one as long as the energy difference
between the levels is smaller than the energy gain due to the
electron--electron interaction.  There is a gain in the interaction
energy because: (i) the direct interaction between the two electrons
in the same state is stronger then the direct interaction between the
electrons in different states; and (ii) electrons occupying different
levels can minimize their interaction energy by maximizing the total
spin, thereby allowing a maximally antisymmetric coordinate wave
function, and gaining the negative exchange energy.

Now, assume that the nanotube initially has $2N-1$ particles where
only the last state is singly occupied so that the total spin is
$1/2$.  The $2N$th electron enters the system at such a gate potential
that the ground state with total spin one is formed.  Then, as the
nonuniform potential is changed further a spin flip occurs and the
spin of the system drops to zero before the $(2N+1)$th electron enters
into the nanotube.  In this scenario both the $2N$th and the
$(2N+1)$th electrons raise the total spin of the system by $1/2$.  But
due to the internal spin flip the total spin of the system is not
larger then one. The entry points of such a scenario are depicted in
Fig.~\ref{fg:levels}b by black triangles.  \vspace{-0.3cm}
\vspace{-0.3cm} We show below that a necessary condition for the
change of the relative positions of the $B$- and $A$- levels is the
presence of a nonuniform gate potential along the nanotube.  In II
Chromium-Gold contacts are deposited on the top of a rope with
nanotubes, and probably break it.  As a result, the only active part
of the nanotube is between the electrodes and is exposed to an almost
uniform gate potential. (Also, there is likely to be strong mixing
between the $A$ and $B$ bands at the damaged tube-ends.)  By contrast,
in I, the nanotube lies on the top of a Si/SiO$_2$ substrate with two
thick Al/Pt electrodes, and it was suggested \cite{NT:Tans97} that the
nanotube is unbroken in that area. The length of the conducting part
of the nanotube ($L\sim 3 \;\mu {\rm m}$) is then larger than the gap
between the electrodes ($d\sim 0.2 \;\mu {\rm m}$).  The contact
electrodes screen the potential applied to the gate, so that only the
nanotube section above the electrode gap is affected by the gate.
Thus, in I, the external potential due to the gate is not uniform.
Relative motion of the $A$ and $B$ bands is thus possible, which can
lead to spin flips.

We turn now to a description of our model calculation.  The energy due
to the interaction between the electrons is calculated within the
Hartree-Fock theory. We assume that the electrons occupy the single
particle states in the presence of a self-consistent external
potential on the tube $\phi(x)$, where $x$ is the coordinate along the
nanotube. The total ground state energy in this approximation is given
by
\begin{eqnarray} 
&&
E[\{N_{\mu \sigma}\}]= 
\sum_{\mu,\sigma,l} \epsilon_{l\mu}[\phi] \; n_{l\mu \sigma}
- h \sum_{\mu} (N_{\mu\uparrow}-N_{\mu \downarrow}) 
\nonumber \\
&&
\hphantom{E[N}
+\frac{1}{2} U_c \left[\sum_{\mu,\sigma} N_{\mu\sigma} - 
2 N_{ion}-N_0\right]^2 
+ \delta U \sum_{\mu,l}  
n_{l\mu\uparrow} n_{l\mu \downarrow}
\nonumber \\
&&
\hphantom{E[N}
+ J \sum_{\mu, \mu'}
N_{\mu \uparrow}
N_{\mu' \downarrow}, 
\label{eq:E}
\end{eqnarray}
where a discrete level $\epsilon_{l \mu }[\phi]$ corresponds to the
$l$th eigenenergy in the band $\mu=A,B$ of the nanotube in the
presence of the space-dependent potential. (The band energies are
defined such that the Fermi level is zero for a neutral nanotube with
$\phi=0$.)  The occupation of the state $|l \mu \sigma \rangle$, is
$n_{l\mu \sigma}=0$ or $ 1 $ and $N_{\mu \sigma}= \sum_{l} n_{l \mu
  \sigma}$, where $\sigma=\uparrow, \downarrow$.  The term
proportional to magnetic field $h=g \mu_B H/2$ is the Zeeman energy,
and $N_{ion}$ is the number of carbon atoms.  The constant $N_0$ is a
continuous parameter, dependent on the work functions of the nanotube
and the contact electrodes, which we choose so that the chemical
potential is zero in the equilibrium state when the contacts and gate
electrodes are grounded.  The values of the charging energy $U_c$, the
excess interaction of two electrons occupying the same level $\delta
U$, and the exchange parameter $J$ are found by considering the real
electron-electron interaction in the nanotube.

Generically, the interaction contains a long-range part, which
influences mainly $U_c$, and a short-range part due to the local
interaction when two electrons occupy a $p$-orbital of the same carbon
atom.  The amplitude of this Hubbard-like term is approximately $ U_H
\approx 15$ eV \cite{NT:Fulde95}.  The behavior of the long-range part
is sensitive to the screening by the metallic electrodes. The
interaction in the section of the nanotube, which is on the metallic
electrodes is screened by the image charges in the metal and falls off
like $R^2/r^3$ for distances $r$ much larger then the nanotube radius
$R$. We find the following values for the interaction
parameters\cite{NT:Byczuk99a}:
\begin{eqnarray}
 3.94 \;[5.04]  >  &U_c/\Delta& > 1.14 \;[1.34], \nonumber \\
  \delta U /\Delta = 0.11\; [0.22],&~& 
  J/\Delta               \ge 0.22 \; [0.44], 
\end{eqnarray}
for $(10,10)\;[(5,5)]$ armchair nanotubes, respectively, where
$\Delta= \pi \hbar v_F/L$ with $v_F=8.1 \times 10^5 m/s$ is the level
spacing at the Fermi energy in each band. The lower bound on $U_c$ is
for the screened interaction and the upper bound is for the unscreened
interaction.  (Higher order correlation effects between the electrons
may renormalize the parameters of the problem \cite{NT:Byczuk99a}.)
We have checked numerically that the long-range part of the
interaction does not influence $J$ and $\delta U$ significantly and
that the effect of a nonuniform potential, which changes the
self-consistent wave functions, is also negligible for the parameter
regions we consider below.

In order to find the evolution of the levels near the Fermi energy
when the potential on the gate electrode $V_{\rm g}$ is changed, we
denote the self-consistent nonuniform potential on the nanotube by
$\phi(x)$, and the energy of the last occupied level in band $\mu$ by
$\epsilon_\mu$.  In the Thomas-Fermi approximation the total number of
states in band $\mu$ up to energy $\epsilon$ is:
\begin{equation}
\label{eq:N}
{\cal N}_\mu [\epsilon,\phi]= \int_{-\infty}^{\epsilon} d\epsilon'
\int dx \; \rho_\mu \left[\epsilon'+\phi(x)\right]
\end{equation}
where $\rho_\mu(\epsilon)$ is the density of states in band $\mu$.  An
application of a potential $V_{\rm g}$ on the gate electrode changes
both $\phi(x)$ and $\epsilon_\mu$ but does not change the number ${\cal
  N}_\mu (\epsilon_{l \mu},\phi)$ which by definition equals $l$.
Thus for a fixed $l$ we find (by formally expanding ${\cal N}_\mu$ with
respect to both its arguments)
\begin{equation}
\label{eq:dEdV}
\frac{\partial\epsilon_{l \mu}}{\partial V_{\rm g}} \approx -
\frac{\int dx \rho_\mu[\epsilon_\mu + \phi_0(x)] \partial \phi(x)/\partial V_{\rm g}}
{\int dx \rho_\mu[\epsilon_\mu + \phi_0(x)]}
\end{equation}
where $\phi_0(x)$ is the self-consistent potential on the tube for
$V_{\rm g}=0$. To model the situation in I we take $\phi_0(x)$ to be a
constant $\phi_0=0.3V$ in the regions on top of the contacts but zero
over the gap, while $\partial \phi(x)/\partial V_{\rm g}$ is equal to
a constant $c$ over the gap and zero in the regions over the contacts.
For the tight-binding-band approximation to an armchair nanotube we
may write: $\rho_\mu(\epsilon)\propto 1/\sqrt{1-[(\epsilon -
  \epsilon_{\mu}^0)/(2\gamma)]^2}$ where
$\epsilon_{A}^0=-\epsilon_B^0=\gamma$ is the position of $\mu$-band
center, and $\gamma=2.7$ eV is a quarter of the width of the
(extended) bands in the nanotube. Thus for $\phi_0 \ll \gamma$ we
find:
\begin{equation}
\label{eq:dEdV1}
\frac{\partial\epsilon_{l \mu}}{\partial V_{\rm g}} \approx - 
c \frac{d}{L} \left[1 \pm \left(1-\frac{d}{L}\right) \frac{\phi_0}{3 \gamma}
 \right], 
\end{equation} 
where $+[-]$ refers to the $A [B]$ band. The value of the constant $c$
is mainly determined by the ratio of $d$ to the distance of the
nanotube from the gate electrode.  Based on a crude model of the
geometry in I, we estimate that for $V_{\rm g}=\pm 4V$ the potential
on the tube is of order $\pm 0.1V$ which gives $c \approx 0.02$
\cite{NT:Byczuk99a}.

Using the picture described above, we find that $\epsilon_{l\mu}
[V_{\rm g}]$ in Eq.~(\ref{eq:E}) is given up to additive constants by
\begin{equation}
\label{eq:single_levels}
\epsilon_{la}[\phi] = l \Delta  - c V_{\rm g}
 ({\alpha}+{\beta}),\;\;
\epsilon_{lb}[\phi] = l \Delta  - c V_{\rm g}
{\beta}, 
\end{equation}
where $ \alpha \approx 0.005$ and $ \beta \approx 0.065$ for
$d/L=1/15$.  The slope of the levels from $A$-band (dashed lines in
Fig.~\ref{fg:levels}a) is $ \alpha + \beta$ and the slope of the
levels from $B$-band (dashed-doted lines) is $ \beta$.

Having crude estimations for the parameters of the model (\ref{eq:E})
we can find a set of conditions under which a sequence of consecutive
electrons will enter into the nanotube with PS.  Fig.~\ref{fg:levels}b
shows the evolution of the ground state spin with a fixed number of
particles when the gate potential increases.  The width $\Delta V$
where the ground state has spin one depends on the values of $\alpha$
and $\beta$ as well as on the interaction parameters $\delta U$ and
$J$.

\begin{figure}[h]
  \vglue 0cm \epsfxsize=1\hsize
  \epsffile{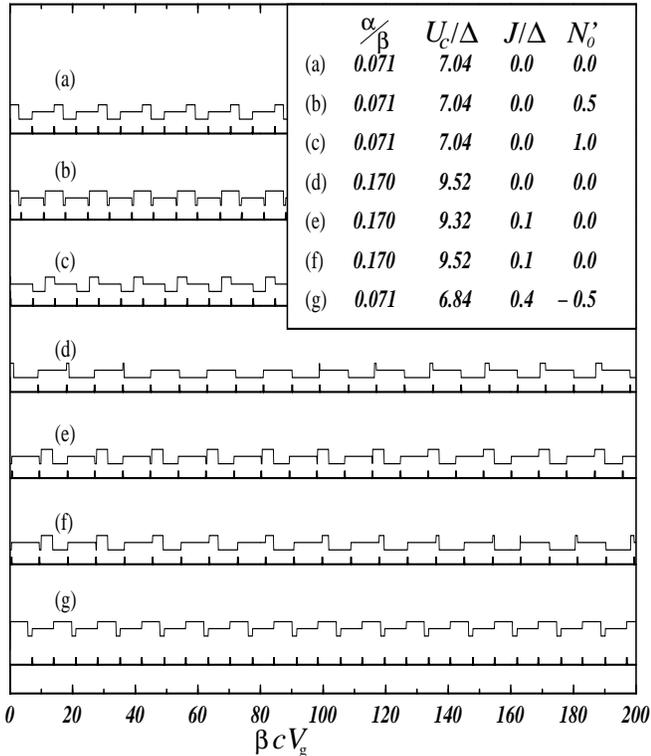}
\caption{\label{fg:lcbs} 
  Evolution of the total spin $S$, under a non-uniform potential,
  whose maximum value on the nanotube is $c V_{\rm g}$.  Various
  cases are shown, with $\delta U =0.2 \Delta$ and with other
  parameters as given in the inset.  The total spin $S=0$, $1/2$ or
  $1$, is shown by the thin solid lines. Tick marks on the thick
  horizontal lines show the entry points of electrons.  The $x$-axis
  is the product $\beta c V_{\rm g} $, in units of the level
  spacing $\Delta$.  A small magnetic field $h=0.001 \Delta$ was
  included. }
\end{figure}

In order to find an example of sufficient conditions to have a PS
sequence when electrons enter into the nanotube we first set $J=0$ and
$\delta U \ll \Delta$.  For a fixed even number $2N$ of particles we
find, after analyzing the evolution of the highest occupied level
(bold line in Fig.~\ref{fg:levels}a), that a spin flip occurs whenever
an empty $A$-level crosses an occupied $B$-level. These crossing
points are marked by $F1,F2,F3$.  Notice that if point $F1$ occurs at
a value $ c V_{\rm g} = 0$, then $f1$ and $N1$ occur at $ c V_{\rm g}
= \Delta/ \alpha $, point $F2$ occurs at $ c V_{\rm g} = 2 \Delta/
\alpha $, etc.  We find that the following conditions are sufficient
for a PS sequence: (i) $N_0$ is such that the initial Fermi energy is
near the crossing point $F1$; (ii) the $(2N+2)$th electron enters at
$\delta V_{\rm g}$ corresponding to points $f1,f2,...$ at which spin
flips occur in the case of $2N+2$ electrons in the system; and (iii)
the $(2N+1)$th electron does not enters at $c V_{\rm g}$ corresponding
to the points $F2,F3,...$.  We note that it is possible to find other
sufficient conditions for a PS sequence, here we give one example.
Since, with the assumptions $J=0$ and $\delta U \ll \Delta$, electron
entries occur when the distance of the line $F1 \rightarrow f1
\rightarrow F2 \rightarrow f2 \dots$ from the Fermi level is equal to
the charging energy $U_c$, we find that (i), (ii) and (iii) are
fulfilled when
\begin{eqnarray}
\label{eq:conditions}
&(i)&\; N_0' \equiv \left(N_0 - k\right) \, {\rm mod} \, 4=0, \nonumber\\
&(ii)& \; U_c^{\ast} /\Delta = 
(1/2) \left[n+ (\beta/ \alpha) 
(2n+1) \right], \nonumber  \\
&(iii)&\;  U_c^{\ast} /\Delta \not = m(1 +  2 \beta/ 
 \alpha) . 
\end{eqnarray}
where $n$, and $m$ and are integers, while $k$ is a large real number,
proportional to $L$, whose value depends on details of the system.
Since $N_0'$ is very sensitive to details, it may be considered a
random number, different for every nanotube.  If $ \alpha / \beta$ and
$N_0$ satisfy the conditions (i)-(iii) we obtain an infinite sequence
of PS electron entries.  Conditions (i) and (ii) may be slightly
relaxed when a finite $\delta U$ is included; in order to find a
finite PS sequence in that case the actual $U_c$ in Eq.(\ref{eq:E})
has to be in a vicinity $\delta U$ of $U_c^{\ast}$ given in
Eq.~(\ref{eq:conditions}), and condition (i) needs to be satisfied
with an accuracy of order $ \delta U/ \Delta$.  However, finite
$\delta U$ makes the condition (iii) somewhat more difficult to
fulfill.  Even with the inclusion of a finite $J$ it is possible to
find, as shown in Fig.~\ref{fg:lcbs}g, a set of parameters that give a
PS sequence.

\vspace{-0.3cm} To illustrate how the PS scenario works, we have
calculated numerically the addition spectrum and the total spin of the
ground states as a function of the maximum voltage on the nanotube, $c
V_{\rm g}$, that varies with the potential on the gate electrode.
Results are shown in Fig.~\ref{fg:lcbs} for few selected sets of
parameters, which are listed in the insert.  Case~(a) corresponds to
an ideal situation when all electrons enter with the same (up) spin;
in case~(c) all the electrons enter with spin down.  In case~(b) $N_0$
is in-between the ideal matching conditions and the spin sequence is
alternating.  These three examples have $J=0$ and parameters
corresponding to $n=0$ in condition (\ref{eq:conditions})ii.
Cases~(d)-(f) correspond to $n=1$. We see that by changing slightly
the model parameters we can get PS series, alternating spins, and
transitions between them.  Finally in case~(g) we show that even with
finite $J$ it is possible to get a PS sequence.

We note that by increasing the value of $h$ by an amount of order
$\Delta$, one can convert a sequence of up spin entries, as in (a), to
a sequence of down-spin entries as in (c)\cite{NT:Byczuk99a}.  This is
consistent with experimental results in I.

In Figure 1, of I, internal transitions of the nanotube appear as
discontinuities of slope in the boundaries of the regions in the plane
of bias voltage and gate voltage where conductance is inhibited by a
Coulomb barrier.  In our model, the relative change in slope on
transition from an $A$-level to a $B$-level should be $\alpha /
\beta$, or about 7\%.  In I, however, the changes observed are much
larger than this.  Such large changes may be more easily explained if
there are localized states at the Fermi energy, due to disorder, which
could then respond quite differently to the gate or bias
voltages\cite{NT:Wingreen99Balents99}.  On the other hand, slope
changes observed in Ref.~\onlinecite{NT:Postma99}, for a different
nanotube, were smaller, of order 20\%, and might possibly be
consistent with our model of a perfect nanotube.

In conclusion, we presented a model of the armchair carbon nanotube in
a nonuniform external potential and elucidated a microscopic mechanism
for internal transitions in which spin flips occur.  This mechanism
could lead to electrons entering in parallel spin sequences, as was
observed in I,  and is consistent with the lack of these 
in II. We note that a sequence of parallel spin entries occur when few
special conditions are fulfilled, in most cases one or more are not
fulfilled and there is alternate or irregular spin sequence.  However,
there is a finite probability, that could be increased with the proper
choice of gate configurations (that control the model parameters), to
find a long sequence of parallel spins entries.

We highly acknowledge discussions with A.~Bezryadin, 
M.~Bockrath, C.~Dekker, C.~M.~Lieber,
H.~Park and T.~H.~Oosterkamp.  This work was supported in part by the
NSF through the Harvard MRSEC (grant DMR 98-09363), and by grants DMR
94-16910, DMR 96-30064, DMR 97-14725.  KB was supported by the
Foundation for Polish Science (FNP).  


\begin{thebibliography}{10}
\bibitem{NT:Remark99a}
Entering of consecutive electrons with parallel spins could be explained if the
  ground state spin of the nanotube was large, but it can be ruled out because
  (i) bulk graphite layers are not spin polarized and (ii) it does not give an
  explanation for the different in the experimental results of I and II.
\end{thebibliography}
\end{document}